ORIGINAL ARTICLE

# Numerical analysis of electron density and response time delay during solar flares in mid-latitudinal lower ionosphere

**Sayak Chakraborty[1] · Tamal Basak[1]**

**Abstract** Impacts of solar flare vary at different parts of the lower ionosphere depending on it's proximity to the direct exposure of incoming solar radiation. The quantitative analysis of this phenomena can be attributed to 'solar zenith angle ($\chi(t)$)' profile over ionosphere. We numerically solve the 'electron continuity equation' to obtain the lower ionospheric electron density profile ($N_e(t)$). The electron production rate ($q(t)$) is governed by the (i) X-ray profile ($\phi(t)$) of the flare, (ii) $\chi(t)$-values during the flare occurrence etc. For analyzing the X-ray profile during flares, we use the GOES-15 satellite observations. Since we're working on electron continuity equation based simplified ionospheric model, we confined our analysis for comparatively stable mid-latitude ionosphere only. We choose three flares each from C, M and X-classes for $N_e(t)$-profile computation. We observe that temporal $N_e(t)$-profiles differ when computed for lower ionosphere over different discrete latitudes. Further, we compute the spatial $N_e(t)$-profile across mid-latitude at the time when $\phi(t) = \phi_{max}$. Now we assume that, these flares repeat themselves every day of a year ($DoY$) at the same time of a day and we compute $N_e(t)$-profiles for each day. We found a seasonal effect on $N_e(t)$-profile due to solar flare. Further, we investigate the response time delay ($\Delta t$) of the lower ionosphere, which is the time difference between incidence of X-ray and the respective change in $N_e(t)$-profiles during solar flares. Strong seasonal effects on $N_e(t)$-profile and $\Delta t$ are the unique results of this work.

✉ T. Basak
tamalbasak@gmail.com

[1] Amity University Kolkata, Major Arterial Road, Action Area II, Rajarhat, New Town, Kolkata 700135, India

## 1 Introduction

Solar-ionosphere interaction during solar flares causes sudden enhancements and thereafter a gradual decay of the ionospheric electron density ($N_e(t)$). Since, the degree of ionization varies widely from layer to layer, it causes a wide variation in the altitude profile of $N_e(t)$. The impact of a solar flare is different at different layers. Le et al. (2012) reported that the Extreme Ultra Violet (EUV) response of a flare is better correlated than X-ray response with Total Electron Content (TEC) of the ionosphere. From SROSS-C2 satellite data analysis, Sharma et al. (2004) reported that the temperature enhancement of ions and electron of ionosphere take place during a flare. Berdermann et al. (2018) reported an overall ionospheric response to an X9.3-class flare and it's consequences on navigation systems. The systematic enhancement of lower ionospheric parameters during a set of solar flares is reported by McRae and Thomson (2004), Pacini and Raulin (2006), Todoroki et al. (2007), Grubor et al. (2008), Singh et al. (2014), Pandey et al. (2015), Selvakumaran et al. (2015), Kumar and Kumar (2018), Chakrabarti et al. (2018), Gavrilov et al. (2019). In this work, we concentrate on the electron density variation of the lower ionosphere (namely, the D-region of ionosphere) only. Under some necessary assumptions, the $N_e(t)$-profile is governed by the famous 'electron continuity equation' (Eq. (1)) (Rowe et al. 1970). In this work, we use this equation and try to obtain the lower ionospheric electron density profile ($N_e$) during the solar flare (Fig. 1). The lower part of the ionosphere consists of almost six thousand types of chemical components with varying abundances. During a solar flare, thousands of complicated chemical processes related to ionization and recombination go on within the lower ionosphere. Those different chemical reactions have widely varying reaction time scales. As a result, the lower ionosphere generally does not respond instantaneously to the incoming solar irradiation, but an effective response time delay ($\Delta t$) comes into the picture because of the collective effects of these respective time scales (Valnicek and Ranzinger 1972; Basak and Chakrabarti 2013). Following Zigman et al. (2007), Basak and Chakrabarti (2013), Palit et al. (2015), we define an effective response time delay ($\Delta t$) as the time gap between the timings of peak X-ray flux ($\phi_{max}$) and peak electron density flux ($N_{e,max}$) during a flare. We computed the '$\Delta t$' from $\phi_{max}$ data we used in this analysis, and from $N_{e,max}$ as computed from 'electron continuity equation'. They reported that $\Delta t$ decreases with increasing $\phi_{max}$ or $N_{e,max}$. Among them, Basak and Chakrabarti (2013) computed the $\Delta t$ from numerical modeling of the sub-ionospheric radio signal propagation effects. Since, the amplitude and phase modulation of the propagating signal through ionosphere goes hand in hand with $N_{e,max}$ profile variation (Zigman et al. 2007), it is justified to do the first order validation of our results with that of Basak and Chakrabarti (2013). Thomson and Clilverd (2001), Le et al. (2007), Basak and Chakrabarti (2013) reported that solar zenith angle ($\chi(t)$) has very crucial role to play to influence the $N_e(t)$. Basak and Chakrabarti (2013) also reported that similar solar flares occurred at different times of a day leaves different levels of imprint on the ionosphere. To investigate this matter, we incorporated a $\chi(t)$ dependent electron production rate ($q(t)$) to the Eq. (1). The $\chi(t)$ is a function of time of a day (UT), latitude/longitude, day of the year ($DoY$) etc. Thus, we're able to generate specific electron production rates ($q(t)$) and supply to Eq. (1).

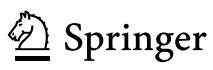

We've restricted our analysis for mid-latitude ionosphere only, i.e., from 30° to 60° latitudes for both northern and southern hemispheres. We purposefully excluded the equatorial and polar ionosphere from this analysis because of the following reasons. (i) The 'plasma bubble effect' and other nonlinear irregularities in the low-latitude or equatorial ionosphere, such as, equatorial ionization anomaly (EIA) and involved equatorial plasma fountain (EPF) (Balan et al. 2018; Carter et al. 2016) are out the scope of computation of our analysis. The particle precipitation effects are responsible for additional modification of the polar ionosphere (Vontrat-Reberac et al. 2001). These types of effects are also not included in our analysis. We computed temporal $N_e(t)$-profile and $\Delta t$ in the lower ionosphere over six different latitudes, namely, 30°N, 45°N, 60°N, 30°S, 45°S and 60°S. We call them as 30N, 45N, 60N, 30S, 45S and 60S respectively from now onwards. For this particular analysis, we choose three different classes of flares, namely, X4.9, M5.2 and C5.2. The details are given in Table 1. The peak region of solar cycle-24 was in the year 2014. A lot of prominent X and M-classes of solar flares occurred around the year 2014. So, it was convenient to choose a set of X and M-classes of flares from there. The X-ray solar flare data ($\phi(t)$) as recorded by GOES-15 satellite is available from NOAA. The available X-ray data are mainly from two channels, namely, (i) hard X-ray (with wavelength range of 0.05 to 0.4 nanometer (i.e. 3.1 to 24.8 keV)) and (ii) soft X-ray (with wavelength range of 0.1 to 0.8 nanometer (i.e. 1.5 to 12.3 keV)). In this case, we choose the soft X-ray data to use in our methodology. Because, at around the ionospheric height of our interest, solar photons having wavelength around average soft X-ray range do induce photo-chemical reactions and enhance electron production rate ($q(t)$). The solar photons having wavelength greater than that get absorbed in E-region, whereas, the same having wavelength less than that may come down to 60 km for further photo-chemical reactions. This X-ray data has the resolution of 2 seconds. Although, due to some possible technical reasons the resolution is less than 2 seconds in few cases. At a few points, it went down up to 100 seconds almost. The least resolution of 100 seconds is also very small in comparison to the total duration of the solar flares chosen for this analysis. So, it is expected that, this limitation won't affect much the final results.

**Table 1** Table for $\Delta t$ in sec

| Class | Date | Time of $\phi_{max}$ (UT) |
|---|---|---|
| C5.2 | 24.02.2014 | 21:37:07 |
| M5.2 | 04.02.2014 | 04:00:54 |
| X4.9 | 25.02.2014 | 00:50:45 |

The objective of this work is to numerically solve the 'electron continuity equation' to obtain the spatio-temporal profile of $N_e(t)$ at lower ionospheric altitudes. Also, we compute the lower ionospheric response time delay ($\Delta t$) profile. Though, the simulation of lower ionospheric parameters during the solar flare is a well studied subject, we particularly check the $N_e(t)$ and $\Delta t$ profiles for a given solar flare over different geographic latitude/longitude and the seasonal effect on them by checking it's variation with $DoY$. Most of the previous works on ionospheric effects of solar X-ray flares are confined to a particular geographical region but in this case we're considering the entire mid-latitude region. Moreover, we assume that a flare repeats itself everyday of the year and we compute the $N_e(t)$ and $\Delta t$ profiles for each of those days ($DoY$). Finally, we repeat all the aforementioned analysis for one flare each from X, M and C-classes. This wide-range approach makes our work unique from the works already done on similar topics. The obtained results bring in front a lot of significant unrevealed features for example, wide range variations of both $N_e(t)$ and $\Delta t$ with latitude/longitude and $DoY$ for a given solar flare. In the Sect. 2, we have formulated the necessary theoretical frameworks regarding the 'electron continuity equation' by applying justified assumptions with possible explanations. In Sect. 3, we implemented an iterative numerical technique to solve that equation with the help of suitable initial condition. In Sect. 4, we elaborately presented all the $N_e(t)$ and $\Delta t$ related results and necessary descriptions. In the Sect. 5 and last, we concluded with all possible explanations.

## 2 Theoretical understanding

According to the fundamental charge neutrality condition, the time rate of change of $N_e(t)$ because of ionization and various recombination processes is governed by the 'electron continuity equation' (Whitten and Poppoff 1961; Rowe et al. 1970; Basak and Chakrabarti 2013; Palit et al. 2016, 2018; Nina et al. 2018). It is as follows,

$$\frac{dN_e(t)}{dt} = \frac{q(t)}{1+\lambda(t)} - \frac{N_e(t)}{1+\lambda(t)}\frac{d\lambda(t)}{dt} - [\lambda^2(t)\alpha_i + \lambda(t)(\alpha_i+\alpha_D)+\alpha_D]N_e^2(t), \tag{1}$$

where, '$q(t)$' is the electron production rate, '$\lambda(t)$' is the negative ions to electron number density ratio, '$N_e(t)$' is the electron density profile at a given ionospheric altitude, '$\alpha_i$' is the ion-ion recombination coefficient and '$\alpha_D$' is the dissociative recombination coefficient. Different first order ionospheric processes related to free electron at a given height is effectively governed by this equation. By solving it, the temporal profiles of $N_e(t)$ can be obtained at any lower ionospheric height. But before that, we need to know about the roles of the several parameters of the Eq. (1).

### 2.1 Electron production rate ($q(t)$)

Electron production on regular basis in the lower ionospheric region is dominated by the Lyman-$\alpha$ radiation coming from the sun. From Chapman's theory, considering the solar radiation to be perfectly monochromatic with intensity '$I$', the electron production rate is the function of ionospheric altitude ($h$), the number of ions produced by the absorption of unit amount of radiation ($C$) and the solar zenith angle ($\chi(t)$) as mentioned in the earlier section also. Now, following the condition of maximum electron production in the ionospheric layers, we derive the total electron production rate over all possible frequencies of solar radiation responsible for electron production and the expression is

$$q(t) = \frac{C\phi(t)cos\chi(t)}{eH}. \tag{2}$$

Here, the '$H$' is called the 'D-region scale height' and according to Mitra (1992) and (Basak and Chakrabarti 2013). In the expression of $H$, '$T$' is the ionospheric temperature at height '$h$', '$m_{avg}$' is the average mass of the particles or ions present in D-Region of ionosphere and '$g$' is the value of acceleration due to gravity at height '$h$'. The quality of approximation in Eq. (2) also depends on the class of the solar flares. According to the typical solar flare X-ray spectrum for the X-class flares, the proportion of photons pertaining to the energies corresponding to the lower end of the soft X-ray energy band is comparatively greater. The opposite is true for the C-class flares as they are typically hundred times weaker than it's X-class counterpart. So, the fraction of total number of photons incident to that particular ionospheric height respectively during X, M and C-classes of flares participating in ionization mechanism vary moderately. Overall, for all classes of flares, significant part of incoming X-ray flux take part in ionization at 74 km ionospheric altitude and hence, the Eq. (2) is well applicable. It is also evident from the computed electron density ($N_e(t)$) variation.

### 2.2 Negative ions to electron number density ratio ($\lambda(t)$)

As discussed earlier, the ionosphere is made up of electrons and several types of positive, negative and cluster ions. Among which the free electron is taken into consideration for this particular analysis. Since, the roll of negative ions are more prominent in the lower ionospheric D-region than all other uppers regions of the ionosphere, the measurement of relative presence of electrons w.r.t. all other negative type of ions is crucial. Hence, the negative ion to electron density ratio is defined as $\lambda(t)$. According to Mitra (1974), Balachandra Swamy (1991), the negative ions are reduced significantly above 70 km. So, at our height of interest (i.e. 74 km above the surface of the earth), $\lambda(t)$ is significantly low and we can neglect the term containing $\lambda^2(t)$ in Eq. (1).

### 2.3 Recombination coefficients

Along with production of electrons, simultaneous recombination takes place in the ionosphere. There are two types of dominant recombination processes which give crucial contribution in the Eq. (1), namely, the '$\alpha_i$' and '$\alpha_D$'. As we're more interested in the effective contribution from all of those recombination processes, we consider the effective recombination coefficient as, $\alpha_{eff} = \alpha_D + \lambda(\alpha_i + \alpha_D)$ (Zigman et al. 2007; Basak and Chakrabarti 2013). $\alpha_{eff}$ is also a function of '$\phi(t)$'.

We solve Eq. (1) in the next section for computation of $N_e(t)$ during solar flares using some further numerical techniques and with the help of some necessary assumptions.

## 3 Methodology

First, we apply some necessary modifications and approximations on Eq. (1). Since, our objective is to compute the $N_e(t)$-profile of lower ionosphere over different latitudes and for different $DoY$ as well, we start with modeling of '$q(t)$'. The '$\chi(t)$' primarily controls $q(t)$ (Eq. (2)). $\chi(t)$ is a function of time of the day, $DoY$, latitude/longitude etc. Through the appropriate value of $\chi(t)$, we obtain different $q(t)$ at different lower ionospheric sections over respective geographical points for different $DoY$. We took the $\chi(t)$ values from the website '*https://www.esrl.noaa.gov*' and supply to Eq. (2). The value of '$C$' is taken as $1.835 \times 10^{17}$ $J^{-1}$ (Whitten et al. 1965). The universal value of '$e$' is 2.71828. Now, we assume a thermal equilibrium at lower ionospheric altitudes and we take $T = 210$ K from Schmitter (2011) as the equilibrium temperature for the calculation of $H$. Again, we use $m_{avg} = 4.8 \times 10^{-26}$ kg (Mitra 1992; Basak and Chakrabarti 2013) calculation of $H$. The value of acceleration due to gravity at lower ionospheric heights is nearly 0.5% lesser than that at earth surface. Hence, it is justified to use $g = 9.81$ $ms^{-2}$. We take another assumption, where we refrain from calculating $N_e(t)$ for the situations when $\chi(t) \geq 85°$. For such values of $\chi(t)$, the local ionospheric D-region almost disappears and it may be considered as good as night-time ionosphere.

To maintain the equilibrium condition in the ionosphere, effective recombination process also gets enhanced during the solar flare. In order to solve Eq. (1), we need to supply the $\alpha_{eff}$-profile during solar flares to it. After analyzing a bunch of solar flares of various classes occurred in the year 2011, Basak and Chakrabarti (2013), reported a profile of $\alpha_{eff}$ with the respective solar flare flux as follows,

$$\alpha_{eff} = \frac{0.375}{\Delta t(N_{e,max} - cos\chi \frac{\phi_{max} g m_{avg} \Delta t}{\rho e k_b T})}. \tag{3}$$

Where, $\rho$ (= 34 eV) is the amount of energy required to create an electron-ion pair. This range of $\alpha_{eff}$ values is also supported by Whitten et al. (1965), Gledhill (1986), Balachandra Swamy (1991), Nina et al. (2011) and references therein. Equation (3) gives a good quantitative idea of the variation of $\alpha_{eff}$ with $\phi(t)$ during a flare (Fig. 2). We computed $\alpha_{eff}$ from Eq. (3) for each and every values of $\phi(t)$ solve Eq. (1). We understand that, the lower ionospheric region of interest of Basak and Chakrabarti (2013) is not exactly same with that of our analysis. But, we're more interested in the effective recombination mechanisms among all types of dominant ions and electron rather than any individual recombination mechanism. Depending on the structural difference in the ionosphere, the individual recombinations rates may differ from region to region but the variation in effective recombination rate overall is within a justified limit.

As per the investigation by Mitra (1974), Balachandra Swamy (1991), the availability of negative ions gets reduced and that of free electrons gets enhanced with increasing lower ionospheric altitudes. As a result of that, the '$\lambda(t)$' decreases with increasing altitude. As discussed in previous section also, the $\lambda(t)$ values become $\ll 1$ for altitudes above 70 km. Moreover, the variation of $\lambda(t)$ with time during solar flare becomes small, so that the we neglect the term containing $d\lambda(t)/dt$ in the Eq. (1). Now, the simplified version of Eq. (1) becomes,

$$\frac{dN_e(t)}{dt} = q(t) - \alpha_{eff} N_e^2(t). \tag{4}$$

It is worth mentioning that, this simplified form of 'electron continuity equation' is not well applicable for altitudes below 70 km. Because, the values of $\lambda(t)$ vary from 30 to 60 (Gledhill 1986) for altitudes below 70 km. For the rest of the work, we compute $N_e(t)$ at 74 km altitude and hence, we can use Eq. (4) with confidence.

From Eq. (4), we calculate the $N_e(t)$-profile during a solar flare. We follow an iterative method, where, we go on calculating $N_e(t)$ at every instant of time starting from an initial time. First, we need an initial condition to start the numerical iteration, which happens to be a value of $N_e(t)$ at the time instant ($t^0$) just before the initiation of the 'precursor stage' of the solar flare. We can call it '$N_e^0$'. It is justified to take $N_e^0$ from '*International Reference Ionosphere - IRI (2016) with IGRF-13 coefficients*', because at that time the solar flare induced ionospheric perturbation effects are yet to begin and the ionosphere can be assumed to be in it's unperturbed state. Then, we compute $dN_e(t)/dt$ at $t = t^0$ by putting $N_e(t) = N_e^0$ in Eq. (4). At this stage, it is important to mention that the time resolution ($\delta t$) of $N_e(t)$-profile is dependent on that of $\phi(t)$. We obtained the $\phi(t)$ from GOES-15 satellite observation, where the time intervals between any two successive observations from case to case are 2.1015, 2.1016, 2.0000 seconds etc. So, the typical time resolution of available data is slightly greater than 2 seconds. Typically, the duration of a solar flare is in between few hundreds to few thousands of seconds. So, this time resolution ($\delta t$) available for computation is really small compared to the duration of a typical flare of any class. Based on this logic, we assume a linear variation of $N_e(t)$ between any two successive data points of computation. Hence, the expression for $N_e^\kappa$ at $t = t^\kappa$ would be,

$$N_e^\kappa = \frac{dN_e^{\kappa-1}}{dt}\delta t + N_e^{\kappa-1}, \tag{5}$$

where, $\kappa = 1, 2, 3 \ldots \nu$. '$\nu$' is the total number of data points present for a particular observation. Again, we compute $dN_e^\kappa/dt$ at $t = t^\kappa$ from Eq. (4) using $N_e^\kappa$ obtained from Eq. (5). This step by step iterative process goes on for all values of $\kappa$ until the entire temporal $N_e(t)$ profile for the flare is being computed. It is noteworthy that, the computed temporal $N_e(t)$-profile becomes self-sustained after few steps of such iterations after $t = t^0$. Then, it hardly depends on the choice of $N_e^0$. Though, IRI (2016) provides the monthly

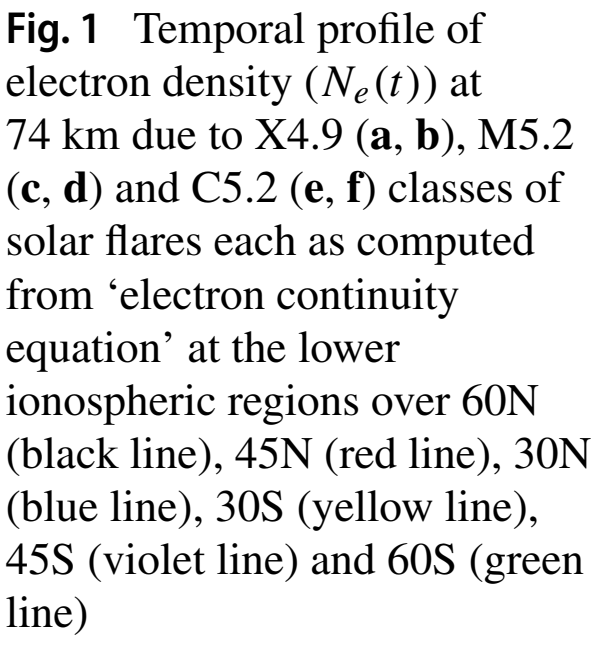

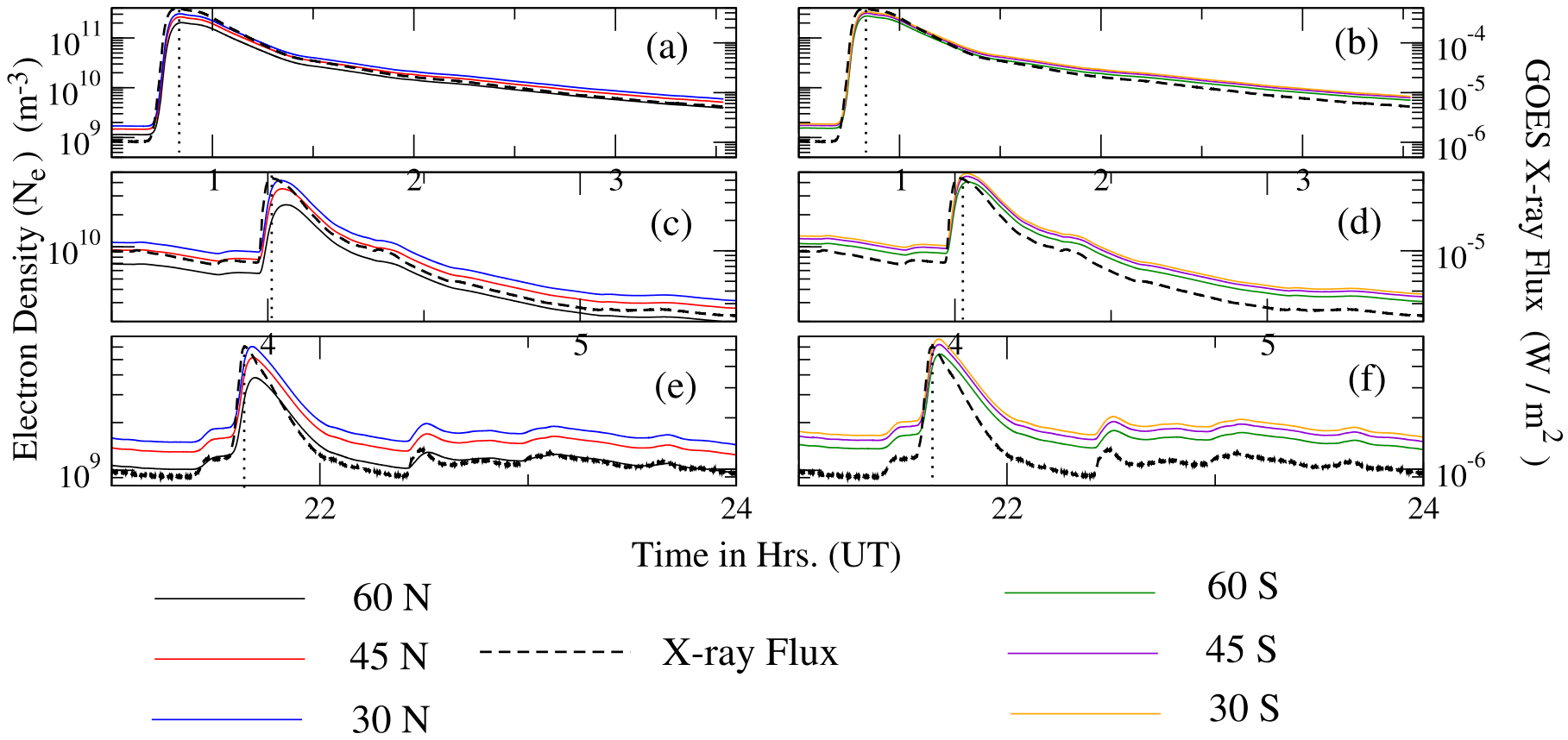

**Fig. 1** Temporal profile of electron density ($N_e(t)$) at 74 km due to X4.9 (**a**, **b**), M5.2 (**c**, **d**) and C5.2 (**e**, **f**) classes of solar flares each as computed from 'electron continuity equation' at the lower ionospheric regions over 60N (black line), 45N (red line), 30N (blue line), 30S (yellow line), 45S (violet line) and 60S (green line)

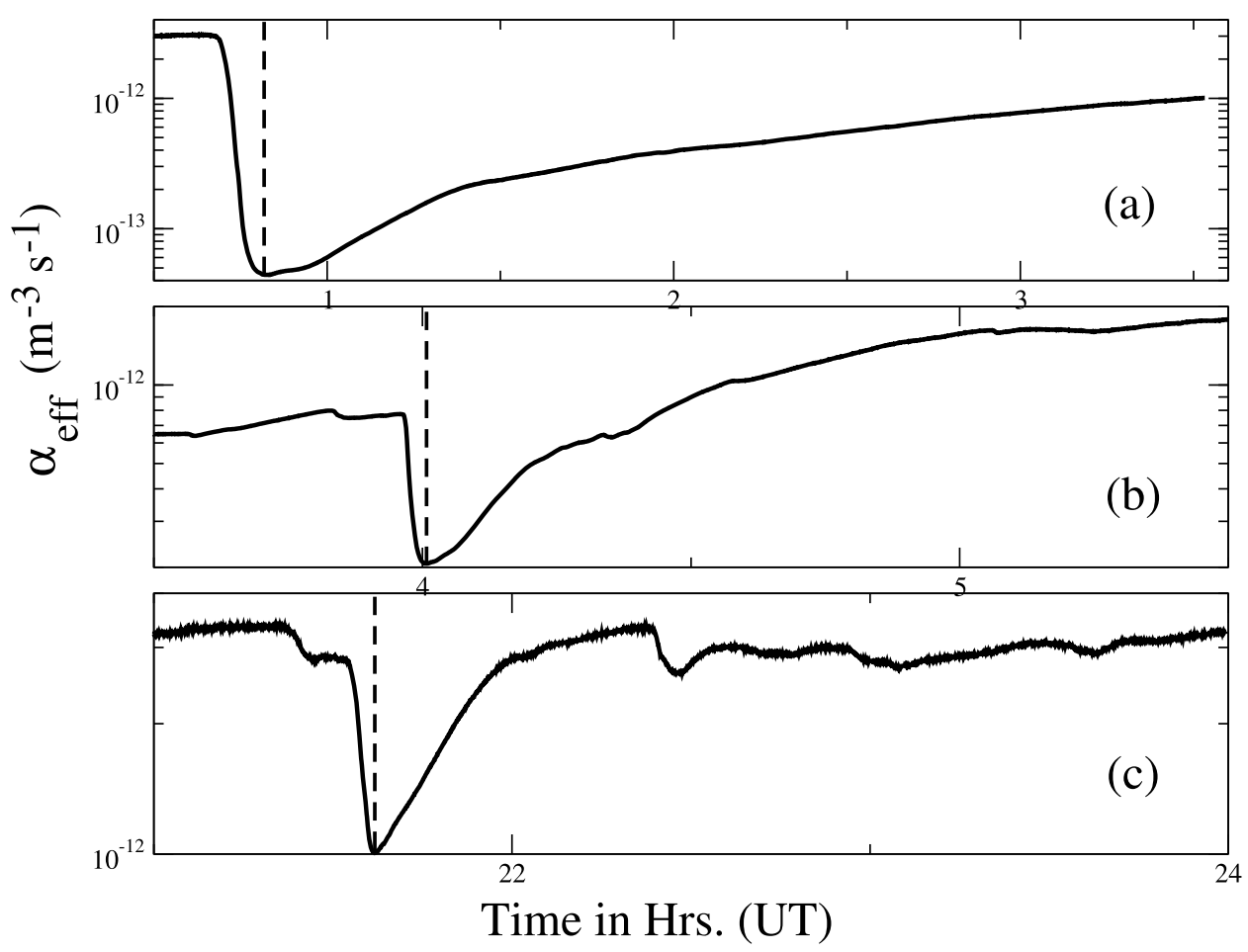

**Fig. 2** Temporal profile of effective recombination coefficient ($\alpha_{eff}$) due to (**a**) X4.9, (**b**) M5.2 and (**c**) C5.2 classes of solar flares each as computed from Eq. (3) at the mid-latitude lower ionospheric regions

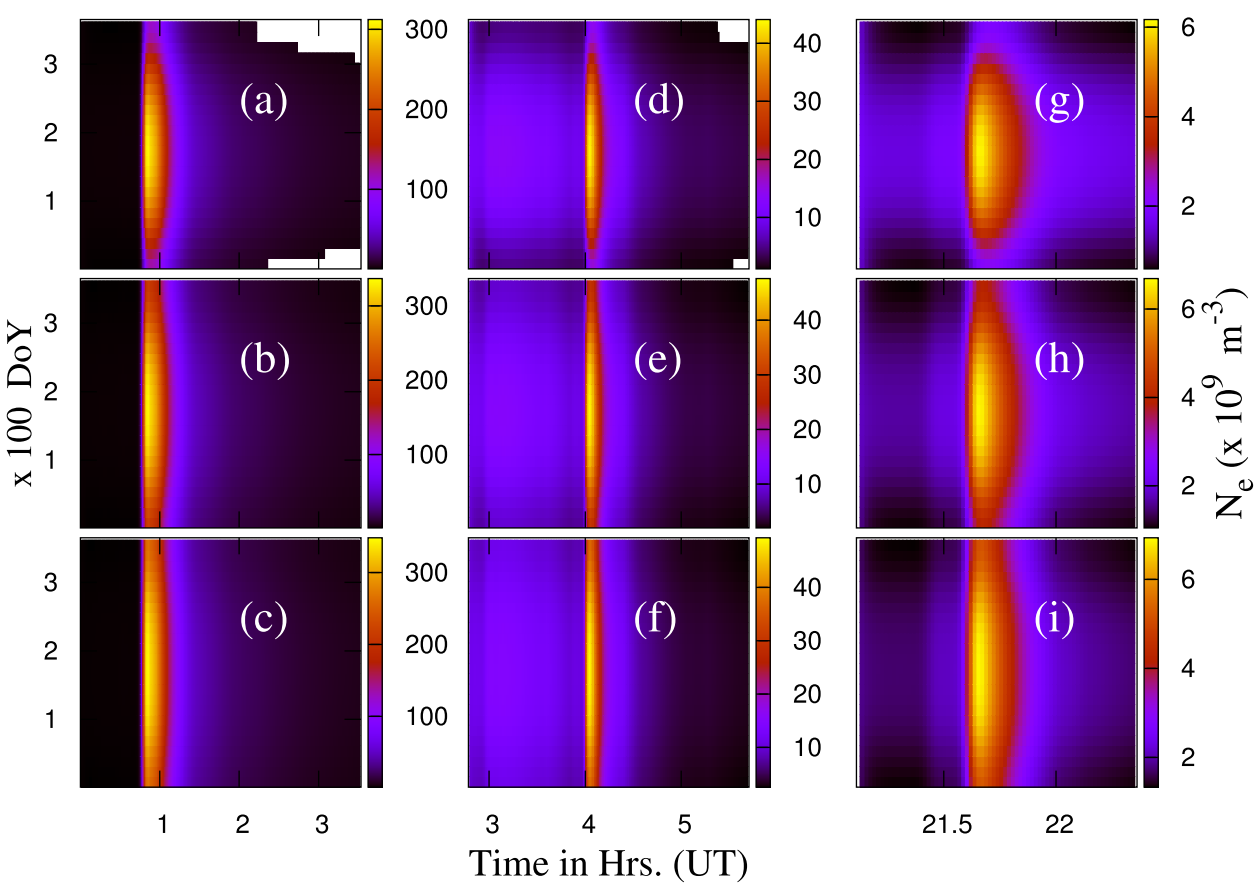

**Fig. 3** Variation of temporal profile of lower ionospheric electron density ($N_e(t)$) with $DoY$ during (**a-c**) X4.9 (**d-f**) M5.2 and (**g-i**) C5.2 classes of flares each at the lower ionospheric regions over 60N, 45N and 30N respectively

average values of $N_e(t)$ and other important ionospheric parameters, it doesn't affect the outcomes of our computation method significantly.

As a natural consequence of $N_e(t)$-profile computation using 'electron continuity equation', we observe that, the computed maximum electron density for a given flare ($N_{e,max}$) does not occur at the same time with respective maximum X-ray brightness ($\phi_{max}$). As we already mentioned that, the time gap between them is known as response time delay ($\Delta t$) of the lower ionosphere. It is expressed as,

$$\Delta t = t_{N_{e,max}} - t_{\phi_{max}} \tag{6}$$

We compute the temporal $N_e(t)$-profile for each of these solar flares (Table 1) at 74 km lower ionospheric altitude over 60N, 45N, 30N, 30S, 45S and 60S respectively (Fig. 1). The chosen longitudes for $N_e(t)$ computation are 167.29E, 119.8E, 144.28W respectively. It was mid-day during the time of $\phi_{max}$ of these solar flares and that is the main reason for choosing those longitude values. In the next step, we compute the temporal $N_e(t)$-profiles in the lower ionosphere over exactly those latitude/longitude assuming the repetition of the flare everyday from $DoY = 1$ to 365 (Figs. 3-4). Now, instead of looking at the $N_e(t)$-profile over some discrete latitude/longitude, we compute spatial $N_e(t)$-profile across entire mid-latitude lower ionosphere at the time when $\phi(t) = \phi_{max}$ (Fig. 5). Finally, we compute $\Delta t$ for each of those flares for all $DoY$s (Fig. 6).

## 4 Result and discussion

In this work, we numerically solved the 'electron continuity equation (Eq. (1))' with the help of Eq. (3) for $\alpha_{eff}$-profile (Fig. 2) and computed the lower ionospheric electron density profile ($N_e(t)$) during three solar flares of X4.9, M5.2 and C5.2-classes respectively (see Table 1). The analysis is divided into two parts mainly. In the first part, we computed temporal $N_e(t)$-profile of the lower ionosphere over different latitudes/longitudes. In the next part, we assumed that exactly these three flares repeat themselves on each day of a year ($DoY$) and computed $N_e(t)$ the $\Delta t$ of the same three flares.

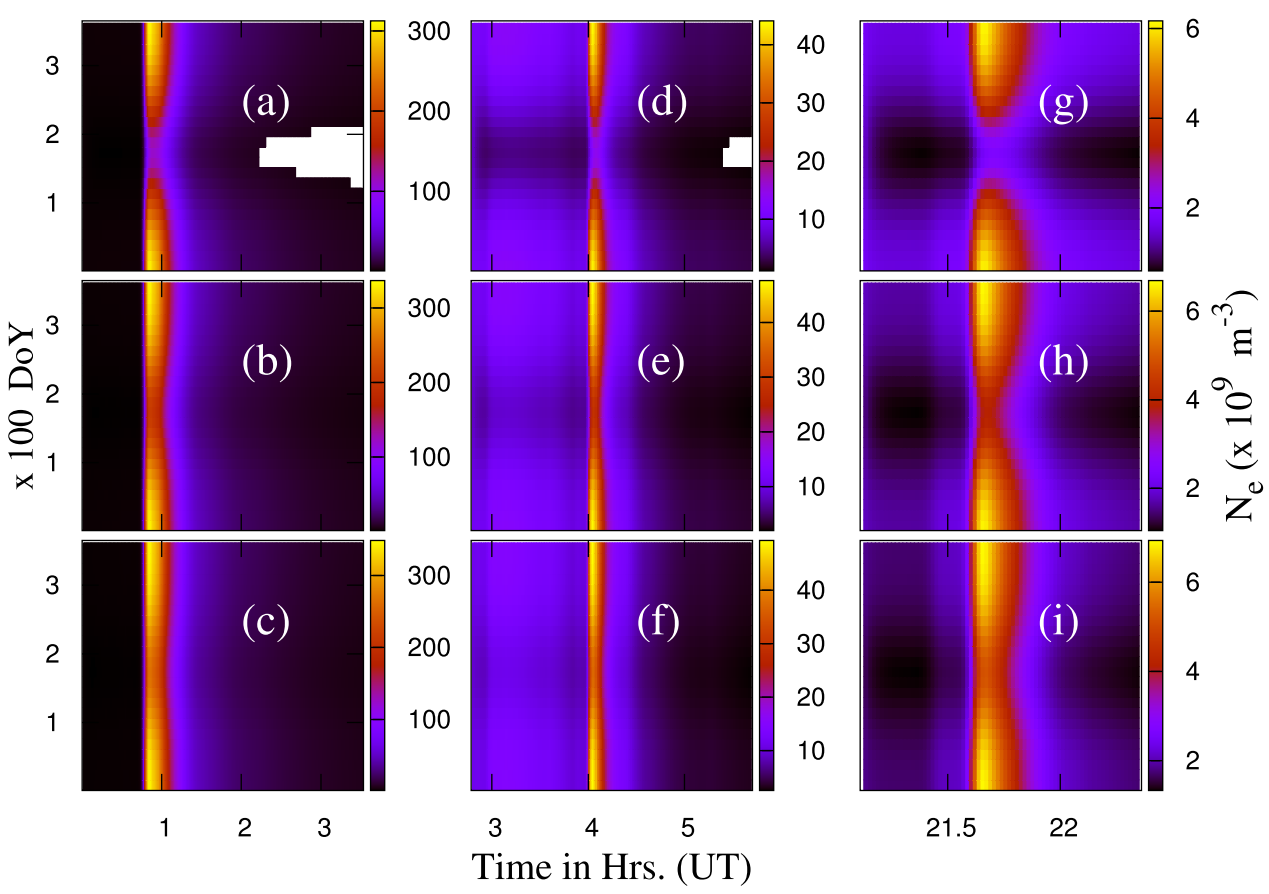

**Fig. 4** Same as Fig. 3 but at the lower ionospheric regions over 60S, 45S and 30S respectively

In the Fig. 1, we choose the specific longitudes, where it was respective mid-days when the maxima of these solar flares were taking place. Since, the solar irradiation is maximum around mid-day, the impact of the solar flares is greatest at that time than any other time of the day. Zigman et al. (2007) reported the $N_{e,max}$s to be around $10^{11}$ m$^{-3}$ and $2.5 \times 10^9$ m$^{-3}$ for typical X and C2.7-classes of flares respectively at 74.1 km ionospheric altitude over 45°N latitude nearly. So, the range of the $N_{e,max}$ values computed by our methodology are within the range to established values. Interestingly, we found that the temporal $N_e(t)$-profiles differ in the ionosphere over different latitudes. The $N_{e,max}$ in the ionosphere over 60N are $3.7 \times 10^9$ m$^{-3}$, $2.3 \times 10^{10}$ m$^{-3}$ and $1.8 \times 10^{11}$ m$^{-3}$ for C5.2, M5.2 and X4.9-class flares respectively, while the same over 60S are $5.2 \times 10^9$ m$^{-3}$, $3.8 \times 10^{10}$ m$^{-3}$ and $2.4 \times 10^{11}$ m$^{-3}$ respectively. Again, the $N_{e,max}$s for C5.2-class flare is $5.9 \times 10^9$ m$^{-3}$ and $6.4 \times 10^9$ m$^{-3}$ in ionosphere over 30N and 30S respectively. We note that, the deviation of $N_{e,max}$ from 30N to 60N is much greater than same from 30S to 60S. Since, all the three solar flares were occurred when it was winter in northern hemisphere, the $N_e(t)$ values are higher in general for lower ionosphere over southern hemisphere. During the occurrence of these flares, the solar irradiation was more slanted over northern hemisphere than its southern counterpart. It could be the possible reason for this results and our method is capable to take into account this effect. If we apply our methodology on solar flares occurred on the other half of a year, we expect exactly opposite results.

It is also observed that for all these flares, the $\phi_{max}$ and $N_{e,max}$ don't occur simultaneously and the lower ionospheric response time delay ($\Delta t$) take place. The $\Delta t$ has been defined and explained in the last section (Eq. (6)). According to Basak and Chakrabarti (2013), Palit et al. (2015), the $\Delta t$ varies from few seconds to few minutes for X to C-classes of flares respectively at 74 km ionospheric altitude. The $\Delta t$ for all latitudes are provided in Table 2. So, the $\Delta t$ computed by our method are in the range of established values. Moreover, the $\Delta t$ for a given solar flare differ notably with latitude. It's values are 127 sec and 106 sec in the ionosphere over 60S and 30S respectively for C5.2-class flare. On the other hand, the value are 26 sec and 16 sec at those same latitudes but for the X4.9-class flare. The range of variation of $\Delta t$ is narrower for a stronger solar flare. Also, the $\Delta t$ values in the ionosphere over 60N are 168 sec and 172 sec for C5.2 and M5.2-class of flares respectively. These two flares occurred on $DoY = 55$ and 35 respectively. The effect of $DoY$ is the possible reason for this apparent opposite nature of $\Delta t$ values. In a later section, we investigate $\Delta t$ in greater detail. From Fig. 1 and data from Table 2, we note that the temporal $N_e(t)$-profile is readily influenced by $DoY$. It means, if the same solar flare would have occurred exactly over same latitude/longitude at the same time of another day of the year, the temporal $N_e(t)$-profile would differ. Kolarski and Grubor (2014) reported two solar flares C9.7 and C9.6 occurred on 7 April 2006 ($DoY = 69$) and 7 September 2005 ($DoY = 250$) respectively. The $N_{e,max}$ values caused by them are $8.61 \times 10^8$ m$^{-3}$ and $3.88 \times 10^9$ m$^{-3}$ respectively. One of the possible reasons for the notable difference in $N_{e,max}$ values is the huge difference of $DoY$ of their occurrence. We check the temporal $N_e(t)$-profile (over the same set of latitudes as in Fig. 1) with $DoY$ (using Eqs. (2), (5) and (6)) assuming that, these three solar flares repeat themselves (Figs. 3-4). The temporal $N_e(t)$-profile is plotted in colour scale. The temporal $N_e(t)$-profile is found to follow the seasonal variation of solar irradiation of respective hemispheres closely. The entire $N_e(t)$-profile gets uplifted in the middle of the year than the profiles in the beginning (or the ending) of the year in the ionosphere over northern latitudes (Fig. 3). The scenario is exactly opposite for the same over southern latitudes (Fig. 4). For example, the $N_{e,max}$ is $3.7 \times 10^9$ m$^{-3}$ and $1.8 \times 10^{11}$ m$^{-3}$ for C5.2 ($DoY = 55$) and X1.8-classes ($DoY = 56$) of flares respectively over 60N. But, if these two flares would have taken place in the middle of the year, the expected $N_{e,max}$ values are approximately $6 \times 10^9$ m$^{-3}$ and $3 \times 10^{11}$ m$^{-3}$ respectively over 60N (Figs. 3g and 3a). There are two important observations regarding this particular analysis, (i) the similarity of the temporal $N_e(t)$-profile variation with typical seasonal variation of solar irradiation is consistent for the three solar flares we analyzed here, and (ii) the difference between the temporal $N_e(t)$-profiles computed during summer and winter becomes greater in the ionosphere over higher latitudes. In Figs. 3a, 3d, 4a and 4d, the blank areas indicate that, the $N_e(t)$ hasn't been calculated on those points because of the $\chi(t) \geq 85.0$ condition.

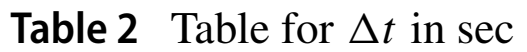

**Table 2** Table for $\Delta t$ in sec

| | 60N | 30N | 60S | 30S |
|---|---|---|---|---|
| C5.2 | 168 | 117 | 127 | 106 |
| M5.2 | 172 | 96 | 93 | 80 |
| X4.9 | 48 | 22 | 26 | 16 |

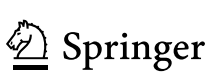

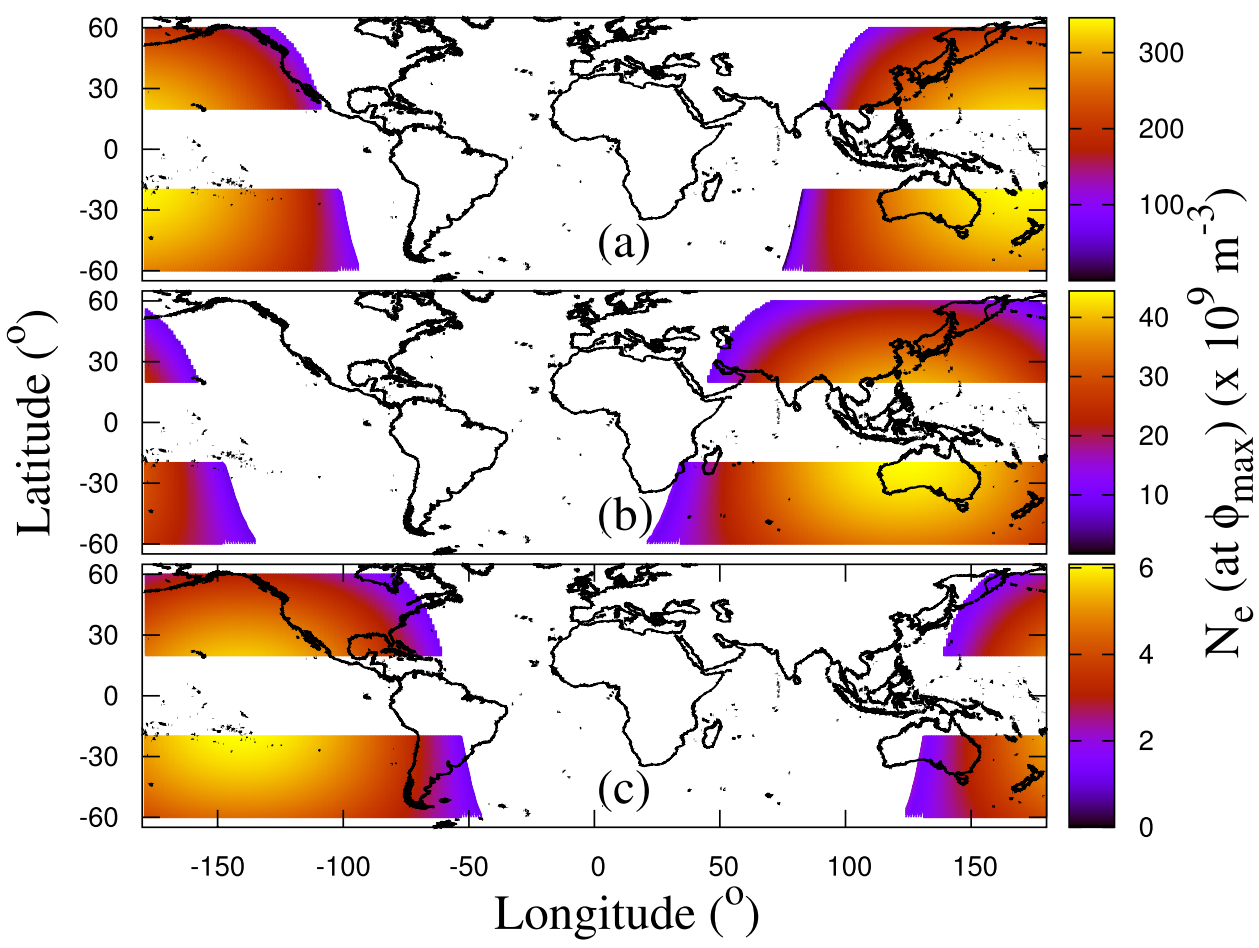

**Fig. 5** Lower ionospheric electron density ($N_e(t)$) across mid-latitude ionosphere at the time of $\phi(t) = \phi_{max}$ for **(a)** X4.9, **(b)** M5.2 and **(c)** C5.2 classes of flares

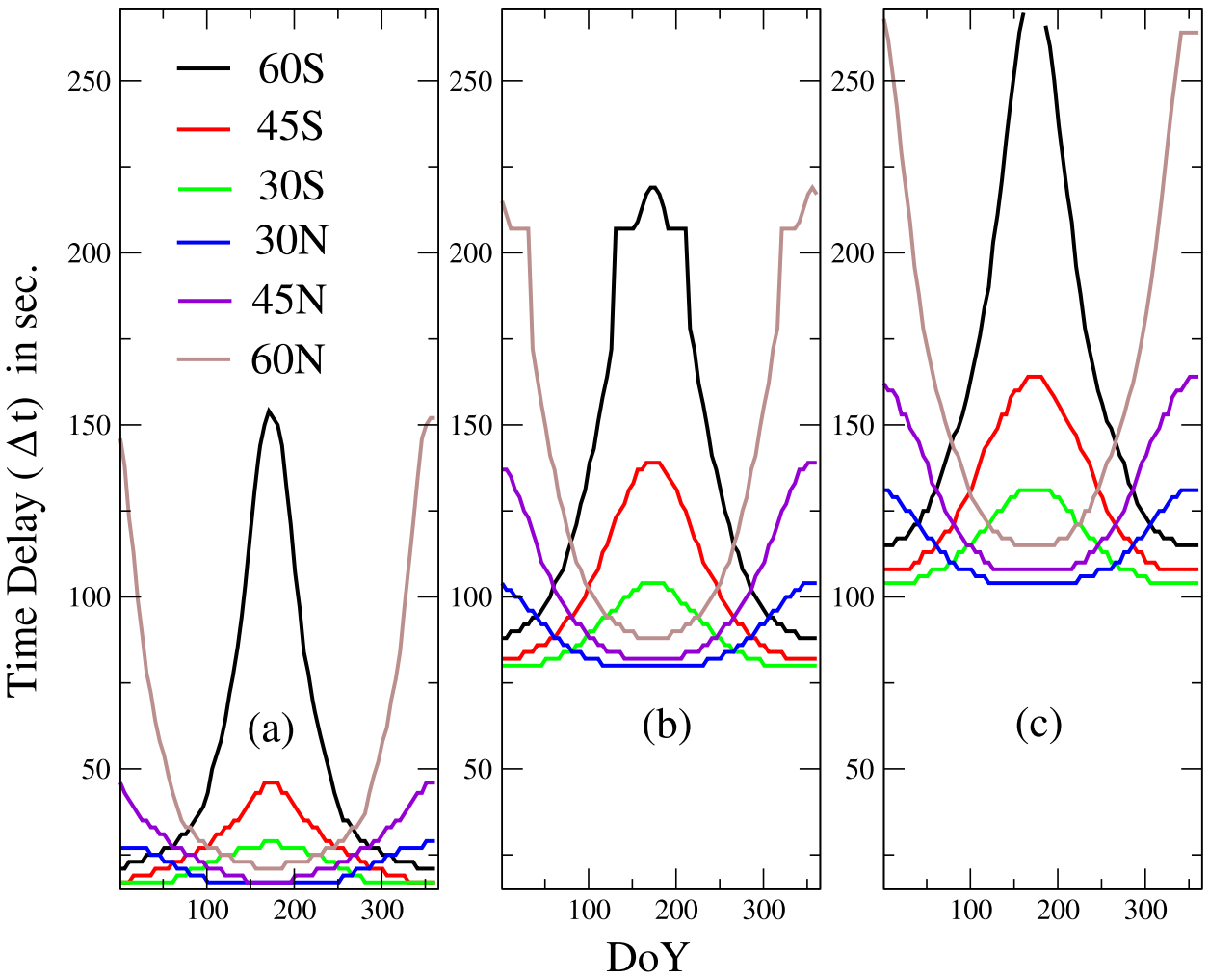

**Fig. 6** Variation of response time delay ($\Delta t$) with $DoY$ during **(a)** X4.9, **(b)** M5.2 and **(c)** C5.2 classes of flares over similar lower ionospheric regions as in Fig. 1

Till now, we computed the temporal $N_e(t)$-profile in the lower ionosphere over some fixed latitudes/longitudes. Now, we concentrate on spatial $N_e(t)$-profile in the mid-latitudinal region of lower ionosphere. We compute the spatial $N_e(t)$-profile for the time when $\phi(t) = \phi_{max}$ (Fig. 5). The $N_e$ (at $\phi_{max}$) is computed only over the illuminated part of the mid-latitude region, i.e., $\chi(t) \leq 85.0$ during $\phi(t) = \phi_{max}$. As a result, the 'day-night terminator' is observed in the spatial $N_e(t)$-profile at $\phi(t) = \phi_{max}$. The $N_e$ (at $\phi_{max}$) values are found to be maximum over 167.29E, 119.8E, and 144.28W longitudes respectively for X4.9, M5.2 and C5.2-classes of flares because around those longitudes, it was mid-day during the occurrence of these solar flares (Figs. 5a, 5b and 5c). The $N_e$ (at $\phi_{max}$) values go down as one moves closer towards 'day-night terminator'. If we take the M5.2-class solar flare as example, the maximum value of $N_e$ (at $\phi_{max}$) is around $4 \times 10^{10}$ m$^{-3}$ and it goes down to $1 \times 10^{10}$ m$^{-3}$ near the terminator (Fig. 5b). Similar fractional change of $N_e$ (at $\phi_{max}$) in the its spatial profile is noted for other two solar flares under consideration (Figs. 5a and 5c). If we do the same operation with a set of similar solar flares occurred in the middle of a year, we expect an inverted profile of $N_e$ (at $\phi_{max}$).

In connection to the characteristics revealed regarding lower ionospheric response time delay ($\Delta t$) from Fig. 1 and Table 2, we follow the same assumption as we did in Figs. 3-4. We compute $\Delta t$ in the lower ionosphere using Eq. (6) over the same set of latitudes as in Fig. 1 (Fig. 6). According to Mitra (1974), Valnicek and Ranzinger (1972), Basak and Chakrabarti (2013), Palit et al. (2015), the ionospheric response is inversely related to the intensity of the ionizing radiation. Since, the solar radiation is much more slanted in nature during winter season, we've the greater values of $\Delta t$. Hence, the $\Delta t$ is greater in winter than in the summer season. That is the reason, it becomes maximum during the middle of the year for south hemispherical latitudes and minimum for north hemispherical latitudes. Like the temporal $N_e(t)$-profiles in Figs. 3-4, this behavior is consistent for these three flares. The difference between maximum and minimum values of $\Delta t$ profile increases with increasing latitude. From Table 2, we've $\Delta t = 106$ sec ($DoY = 55$) and 16 sec ($DoY = 56$) for C5.2 and X4.9-class flares over 30S. If it were occurred at the middle of the year, the expected $\Delta t$ values according to our model are 130 sec and 29 sec respectively. Again, we've $\Delta t = 172$ sec ($DoY = 35$) over 60N for M5.2-class flare. The same flare would have experienced $\Delta t = 88$ sec only if it'd occur in the middle of the year (Fig. 6b). From $DoY = 163$ to 182, the $\Delta t$ could not be computed for C5.2-class over 60S (Fig. 6c), because the ionosphere was in night-time condition before the solar flare effect could be completed. The variations of $\Delta t$ for M5.2-class flare especially over 60S and 60N (Fig. 6b) aren't as smooth as the same for X4.9-class flare (Fig. 6a). There is a step function like behavior in Fig. 6b for 60S and 60N. Hence, piecewise constant value of $\Delta t$ is observed. In the middle of the year over 60S and at the end/beginning of the year over 60N, the '$\chi(t)$' values are quite high and as a result $\Delta t$ values purely due to the M5.2-class flare couldn't be reflected. But, for the X4.9 class flare (Fig. 6a), similar effect due to high $\chi(t)$ isn't present because the flare itself is strong enough to makeup the situation.

## 5 Conclusion

We numerically solved the 'electron continuity equation (Eq. (1))' and computed the electron density profile ($N_e(t)$) of lower ionosphere (namely, the D-region of ionosphere) during three solar flares, namely X4.9, M5.2 and C5.2-classes. Those flares occurred on 25th, 4th, 24th Feb 2014 respectively. We used the solar X-ray profile ($\phi(t)$) from GOES-15 satellite observation. We expressed the lower ionospheric electron production rate ($q$) as a function of global solar zenith angle ($\chi(t)$) profile so that, we became able to incorporate the variation '$q(t)$' specific to latitude/longitude and day of the year ($DoY$) to Eq. (1). As a result, we obtained $N_e(t)$-profile and response time delay ($\Delta t$) specific to the lower ionosphere over particular latitude/longitude and $DoY$. Though, the lower ionospheric $N_e(t)$-profile (Rowe et al. 1970; Thomson and Clilverd 2001; Zigman et al. 2007; Nina et al. 2011; Basak and Chakrabarti 2013) and response time delay ($\Delta t$) (Valnicek and Ranzinger 1972; Mitra 1974; Basak and Chakrabarti 2013; Palit et al. 2015) are well studied subject, but this latitude/longitude and $DoY$ specific results make it unique. The profile of $\Delta t$ as presented here is a natural outcome of the solutions of 'electron continuity equation (Eq. (1))' and it is consistent with the established values in the literature. Here, we mention that, we've used an approximate $\alpha_{eff}$ profile (Fig. 2) to solve Eq. (4) instead of incorporating a detail $\alpha_{eff}$ profile suitable for mid-latitude lower ionosphere. Despite it, our methodology proves itself capable of producing such detail results.

We computed the temporal $N_e(t)$-profile in the ionosphere over six different latitudes, namely, 60N, 45N, 30N, 30S, 45S and 60S. We found that the $N_e(t)$-profile gradually reduces in the ionosphere over higher latitudes. Also, we found that the $N_e(t)$ values are lower over 60N, 45N and 30N than 60S, 45S and 30S respectively. Because, these solar flares occurred in February, when it is winter season in the northern hemisphere. So, we can conclude that the $N_e(t)$ values become lesser when the solar irradiation is more slant. To verify the fact, we compute spatial profile of $N_e(t)$ (at $\phi_{max}$) across entire latitude ranges starting from 30N to 60N and 30S to 60S (Fig. 5). We compute the temporal $N_e(t)$-profile for all $DoY$ assuming that each of these flares repeat itself at the same time of everyday of a year (Figs. 3-4). We report that for all of these flares, $N_e(t)$-profiles gets increased by a factor in the middle of the year for ionosphere over northern latitudes and exactly opposite happens for southern latitudes. Numerical modeling of lower ionospheric electron density variation over a limited geographical zone has been done on several occasions, but a complete mid-latitudinal $N_e(t)$-profile for all three classes of flares using solar zenith angle ($\chi(t)$) as a tool is an unique approach of its kind.

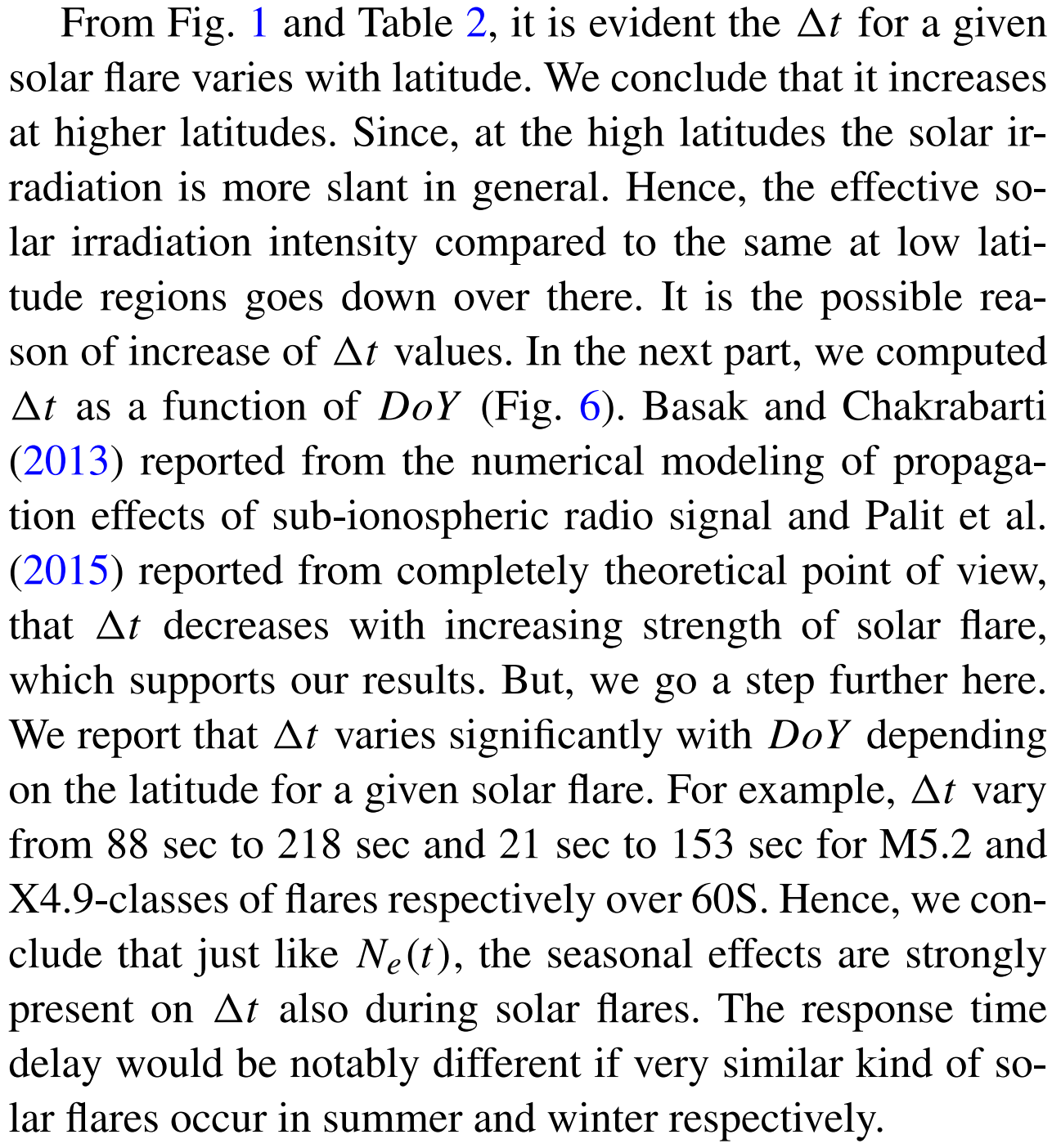

From Fig. 1 and Table 2, it is evident the $\Delta t$ for a given solar flare varies with latitude. We conclude that it increases at higher latitudes. Since, at the high latitudes the solar irradiation is more slant in general. Hence, the effective solar irradiation intensity compared to the same at low latitude regions goes down over there. It is the possible reason of increase of $\Delta t$ values. In the next part, we computed $\Delta t$ as a function of $DoY$ (Fig. 6). Basak and Chakrabarti (2013) reported from the numerical modeling of propagation effects of sub-ionospheric radio signal and Palit et al. (2015) reported from completely theoretical point of view, that $\Delta t$ decreases with increasing strength of solar flare, which supports our results. But, we go a step further here. We report that $\Delta t$ varies significantly with $DoY$ depending on the latitude for a given solar flare. For example, $\Delta t$ vary from 88 sec to 218 sec and 21 sec to 153 sec for M5.2 and X4.9-classes of flares respectively over 60S. Hence, we conclude that just like $N_e(t)$, the seasonal effects are strongly present on $\Delta t$ also during solar flares. The response time delay would be notably different if very similar kind of solar flares occur in summer and winter respectively.

**Acknowledgements** Authors acknowledge NOAA's National Centers for Environmental Information (NCEI) for using solar X-ray data and International Reference Ionosphere (IRI) (an international project sponsored by the Committee on Space Research (COSPAR) and the International Union of Radio Science (URSI)) for using electron density data.